\documentclass[twocolumn]{aastex631}

\def\fedd{{\rm f}_{\rm Edd}}
\newcolumntype{C}{>{$}c<{$}}
\usepackage{amsmath}

\newcommand{\MSun}{{\rm M}_{\sun}}
\newcommand{\dorb}{\delta \phi_{\rm orb}}
\newcommand{\decc}{\delta \phi_{\rm ecc}}
\newcommand{\dGW}{\delta\phi_{\rm GW}}

\begin{document}

\title{Gas-induced perturbations on the gravitational wave in-spiral of live post-Newtonian LISA massive black hole binaries}

\correspondingauthor{Mudit Garg}
\email{mudit.garg@uzh.ch}

\author[0000-0002-9032-9103]{Mudit Garg}
\affiliation{Department of Astrophysics, University of Zurich, Winterthurerstrasse 190, CH-8057 Z\"urich, Switzerland}

\author[0000-0002-8400-0969]{Alessia Franchini}
\affiliation{Department of Astrophysics, University of Zurich, Winterthurerstrasse 190, CH-8057 Z\"urich, Switzerland}

\author[0000-0001-6106-7821]{Alessandro Lupi}
\affiliation{DiSAT, Università degli Studi dell’Insubria, via Valleggio 11, I-22100 Como, Italy}
\affiliation{INFN, Sezione di Milano-Bicocca, Piazza della Scienza 3, I-20126 Milano, Italy}

\author[0000-0001-7889-6810]{Matteo Bonetti}
\affiliation{Dipartimento di Fisica ``G. Occhialini'', Universit\`a degli Studi di Milano-Bicocca, Piazza della Scienza 3, I-20126 Milano, Italy}
\affiliation{INFN, Sezione di Milano-Bicocca, Piazza della Scienza 3, I-20126 Milano, Italy}

\author[0000-0002-7078-2074]{Lucio Mayer}
\affiliation{Department of Astrophysics, University of Zurich, Winterthurerstrasse 190, CH-8057 Z\"urich, Switzerland}

\begin{abstract}
We investigate the effect of dynamically coupling gas torques with gravitational wave (GW) emission during the orbital evolution of an equal-mass massive black hole binary (MBHB). We perform hydrodynamical simulations of eccentric MBHBs with total mass $M=10^6~\MSun$ embedded in a prograde locally isothermal circumbinary disk (CBD). We evolve the binary from $55$ to $49$ Schwarzschild radii separations using up to 2.5 post-Newtonian (PN) corrections to the binary dynamics, which allow us to follow the GW-driven in-spiral. For the first time, we report the measurement of gas torques onto a live binary a few years before the merger, with and without concurrent GW radiation. We also report the gas-induced orbital dephasing $\dorb\sim-0.007$ rad over $278$ orbital cycles that is likely driven mainly by disk-induced precession and LISA should be able to detect it at redshift $z=1$. Our results show how GWs alone can be used to probe the astrophysical properties of CBDs and have important implications for multi-messenger strategies aimed at studying the environments of MBHBs.
\end{abstract}


\section{Introduction} 
\label{S:Intro}

The recent adoption of LISA \citep{AmaroSeoane2017,Colpi2024} and the development of TianQin \citep{Li2025} and Taiji \citep{Gong2021} will provide a powerful opportunity to detect gravitational waves (GWs) from coalescing near-equal mass massive black hole binaries (MBHBs) with masses $\sim 10^4$-$10^7~\MSun$. 
LISA can potentially detect MBHBs up to redshifts $z\lesssim20$ and with high (e.g. $\lesssim10^3$) signal-to-noise ratios \citep[SNRs;][]{AmaroSeoane2017}. 
MBHBs are a by-product of galaxy mergers \citep{Begelman1980}. When two galaxies merge, the massive black holes (MBHs) hosted in their centre are expected to reach the centre of the remnant galaxy owing to the dynamical friction mechanism and form a bound binary at pc scales. This binary can further proceed towards merger through the interaction with surrounding stars and gas until GWs are strong enough to take over and drive the binary to coalescence \citep{AmaroSeoane2023}. 

While interactions with stars in a tri-axial potential \citep[e.g.][]{Quinlan1996,Preto2011,Khan2011} as well as a third close-by MBH \citep{Blaes2002,Hoffman2007,Bonetti2019} can lead to a MBHB merger, they are relatively slow and rare mechanisms, respectively. In particular,
sinking timescales of MBHBs due to three body encounters with stars
can exceed a Gyr in the low density environments of stellar-disk dominated
galaxies \citep{Khan2018}, the typical hosts of MBHs below $10^6~\MSun$, that fall in the mass range accessible by LISA. On the other hand, when the host galaxies are gas-rich, and have circumnuclear gas disks, then MBHBs can sink efficiently below pc separations \citep{Mayer2013,SouzaLima2020}. At separations below $\sim 0.1$ pc, the gas dragged by the two MBHs is expected to settle in a common circumbinary disk (CBD; \citealt{Escala2004,Cuadra2009,DOrazio2016}), whose torques remove angular momentum from the binary allowing it to coalesce in less than $100$ Myr \citep{Haiman2009}. The binary potential will open a cavity inside the CBD whose size depends on the disk properties \citep{artymowicz1994,artymowicz1996}. Over the past few years, the interaction between a binary and its circumbinary disk has been studied extensively for various system parameters and thermodynamics assumptions using different numerical hydrodynamical (HD) simulations \citep[see e.g.][]{Duffell2024}. The general consensus is that circular nearly equal-mass binaries do undergo out-spiral in relatively thick CBDs while their in-spiral is aided by relatively thin (i.e. aspect ratio $H/R \lesssim 0.03$) CBDs \citep{Tiede2020,Franchini2021,Franchini2022}. 

Since MBHBs observable by LISA are likely to reside in gaseous environments \citep[see, e.g.][]{Mangiagli2022}, it is important to study the effect of gas on the orbital evolution of the MBHB when it enters the LISA band.
The first attempt in this direction was performed by several groups \citep{Garg2022,Tiede2024a,Zwick2024,Garg2024b,Garg2024c,Garg2024d,Dittmann2023} who measured the gas-induced dephasing in the LISA band by simply linearly adding the gas-driven evolution rate, computed in post-processing from 2D HD fixed binary orbit simulations, to the GW in-spiral rate. However, the scales considered in those numerical works are close to sub-pc, where GWs are still too weak to drive significant binary evolution. 
Furthermore, by adding the two contribution linearly, gas-induced dephasing studies might have ignored possible coupling between gas torques and GW-driven evolution, due to the lack of HD simulations where the two effects are naturally coupled together and the binary evolves under both processes at the same time. 

Recently, \citet{Franchini2024} simulated an eccentric, live \citep{Franchini2023}, equal-mass $10^6~\MSun$ MBHB embedded in a prograde $100~\MSun$ CBD by dynamically modeling the binary in-spiral with post-Newtonian (PN) corrections up to $2.5$ order. They evolved the system for the final years of in-spiral, including the merger and post-merger phase, to quantify possible electromagnetic (EM) counterparts.
In this work, we use the same setup to simulate the same binary but now embedded in a lighter $5~\MSun$ disk to properly investigate how gas perturbs the binary evolution rate. We then quantify for the first time the effect of gas-induced perturbations on waveforms using a live binary whose dynamics is computed using PN
corrections, thus including the interplay between energy and angular momentum change caused by both GW radiation and gas torques. 
With this simulation setup, the in-spiral is concurrently determined by GWs and gas, as opposed to co-adding the two effects in post-processing  as previously 
done in the literature, allowing us to robustly quantify the gas-induced dephasing in the GW waveform and its detectability by LISA.

\section{Numerical setup} 
\label{S:setup}

Following the approach used in \cite{Franchini2024}, we model the binary using two equal-mass sink particles \citep{Bate1995} that represent two Schwarzschild MBHs with total mass $M=10^6~\MSun$. We set each sink particle radius to the innermost stable circular orbit (ISCO) for a non-spinning MBH. We set the MBHB initial semi-major axis (SMA; $a$) to $a=54.5$ Schwarzschild radii ($r_s$) and eccentricity to $e=0.3$. 
We take the initial SMA to be twice the decoupling radius, theoretically estimated by \cite{ArmitageNatarajan2002}.
Note that we start from the initial condition of the thin (i.e. aspect ratio $H/R=0.03$), locally isothermal disk simulation in \citet{Franchini2024}, which originated from a circular equal-mass binary evolved for $1000$ binary orbits by \citet{Franchini2022}. During the first $1000$ orbits the eccentricity of the simulated live binary increased to $e=0.3$ as a result of the interaction with the CBD. We here assume the disk to have a 
mass $M_{\rm d}=5~\MSun$ {instead of $100~\MSun$ disk in \citet{Franchini2024} to limit the Eddington ratio $\fedd$ at a value closer to unity.\footnote{$\fedd$ is the ratio of the accretion rate onto the binary $\dot M$ and the Eddington accretion rate $\dot M_{\rm Edd}$, which is $M/50$ Myr for our assumed $10\%$ radiative efficiency.}}
The disk is 3 dimensional and initially sampled with $N=4\times 10^6$ gas particles distributed with an initial surface density profile $\Sigma \propto R^{-3/2}$. We use the Shakura-Sunyaev \citep{ShakuraSunyaev1973} turbulent prescription for viscosity, with viscosity coefficient $\alpha = 0.1$, which leads to a kinematic viscosity value $\nu=\alpha c_{\rm s}H = {7.76\times 10^{18}~{\rm cm}^2{\rm s}^{-1}}$ at $R=3a$. The disk equation of state is locally isothermal with the sound speed profile used in \cite{Farris2014}. In the initial setup the disk extended from $2a$ to $10a$. However during the first 1000 binary orbits \citep{Franchini2022}, the cavity becomes eccentric and the inner edge increases to $\sim3.5a$. 

We explore three resolutions for our {gas+PN2.5} setup by increasing the number of splitting levels in the hyper-Lagrangian refinement. We quantify the resolution in terms of inter-particle spacing $\Delta x$ at $R=3a$. We label these simulations low-resolution (LR) with $\Delta x[3a]={1.31~r_s}$, mid-resolution (MR) with $\Delta x[3a]={1.07~r_s}$, and high-resolution (HR) with $\Delta x[3a]={0.66~r_s}$. We run at least $100$ orbits for each simulations in order to perform a meaningful resolution study, which we report in Appendix~\ref{App:Res}. We find that the gravitational torque exerted by the disk onto the binary in the MR simulation is already converged and therefore we consider the MR run as our fiducial setup. Unless stated  otherwise, the following results have been inferred using our MR run.

We follow the evolution of the binary driven by both gas torques and PN corrections up to 2.5PN order using the code {\sc gizmo} \citep{Hopkins2015} until the binary reaches $48.9\,r_{\rm s}$ in separation in our {gas+PN2.5} simulation, i.e. for $278$ initial binary orbits or $\sim556$ GW cycles. The implementation of the PN corrections to the binary dynamics follows the equations in \cite{Blanchet2014}. We include both conservative $1$PN and $2$PN terms, and radiative $2.5$PN terms. 
The latter term generates the GW emission and leads to the decrease in binary SMA ($\dot a_{\rm GW}$) and eccentricity only due to GWs. In order to integrate the 2.5PN equations, we  implemented an intermediate predictor step to update the particle velocities at the end of the time step, accounting for the PN corrections, and re-enforcing the numerical stability of the integration algorithm. Our approach is similar to the one outlined in Sect. 6.2 of \cite{Liptai2019}, except that we use a predictor-corrector approach instead of implementing the implicit kick-drift-kick one (see \cite{Franchini2024} for more details).

In order to measure the effect of the gas contribution in the absence of GW emission, we run a simulation with the CBD, but without the 2.5PN dissipative term in the binary orbital motion. We refer to this simulation with the term ``{gas+PN2}" and we perform it only at mid-resolution, i.e. with $\Delta x[3a]={1.07~r_s}$.
We run this simulation for the same time as the {gas+PN2.5} run to see any appreciable changes in the orbital quantities due to the sole interaction with the gaseous disk. This allows us to infer the gas torques and gas-induced orbital {as well as binary precession} dephasings without the effects introduced by GWs dissipation. Note that we extrapolate the {mean torque} results of the {gas+PN2} simulation {over the same integrated time} down to $48.9~r_{\rm s}$ since evolving the binary under the mere effect of gas to such small separations is currently computationally prohibitive. {Similarly, we compare the binary evolution between different setups over the same integrated time to compute the dephasings.}

We then also run two simulations without the CBD in order to self-consistently obtain the binary evolution driven only by the PN terms. We run one simulation including all the PN corrections, including the dissipative GW term, and another simulation with only the 1+2 PN corrections. We refer to the first and second simulation with the label ``{PN2.5}" and ``{PN2}" respectively. These simulations allow us to isolate the effects of GW emission and of the interaction with the disk, and to mitigate numerical errors in the integration when we compute the difference between the simulation with gas and their non-gaseous counterparts.

\subsection{Post-processing analysis}\label{Sec:Postprocess}

The CBD affects the binary evolution by exerting both a gravitational torque ($T_{\rm grav}$) and an accretion torque ($T_{\rm acc}$). The first, just due to gravity, is essentially driven by any asymmetry in the gaseous flow while the latter is instead induced by the accretion of gas particles onto either MBHs. In particular, the accretion of gas alters not only the mass but also the angular momentum of the binary (see \citealt{Franchini2021} for detailed calculations).

We can then express the overall gas effect in terms of a single dimension-less simulation-calibrated parameter (similar to the accretion eigenvalue mentioned in \citealt{Duffell2024})
\begin{equation}\label{eq:xi}
    \xi=\frac{T_{\rm grav}+T_{\rm acc}}{\dot M a^2\Omega},
\end{equation}
Here $\Omega\equiv\sqrt{GM/a^3}$ is the binary orbital angular frequency, and  $\dot Ma^2\Omega$ is simply the normalization commonly used in the literature \citep[see, e.g.][]{Duffell2024}. $\xi$ depends sensitively on the binary and disk parameters, in particular on the binary mass ratio $q$, disk shape and temperature, and may also depend upon the assumed equation of state. 
Previous 2D Newtonian simulations, featuring sub-pc fixed binary orbit, predicted $|\xi|\lesssim2$ \citep{Dittmann2022} for $H/R\sim0.03$, although for a higher kinematic viscosity value than the one we simulate in this work.

Note that, due to our live binary setup, we can directly compute the change in SMA and eccentricity from the positions and velocities of the binary components, as well as the accretion rate. We can therefore measure $\xi$ and $\fedd$ independently for both {gas+PN2.5} and {gas+PN2} simulations. 

The inclusion of gas in the MBHB system induces perturbations in the evolution of the binary orbital phase. The difference in the number of binary cycles before the merger when it evolves in a gaseous environment is the result of a combination of environmental effects. Indeed the gas causes a different evolution of the binary semi-major axis, eccentricity and precession rate.
Since we lack analytical prescriptions for each of these terms in the PN approximation, we can only directly measure the gas-induced orbital ($\dorb$) and precession ($\decc$) dephasings by comparing the simulations with and without the gaseous disk.
We therefore compare the {gas+PN2.5} and {PN2.5} orbital phases to infer the global effect that the gaseous disk has on the number of binary cycles {of a GW-driven MBHB}.
We then compare the {gas+PN2} and the {PN2} runs to better isolate the effect of gas on the binary precession rate without any perturbations to the binary induced by GW emission. Indeed in {the gas+PN2 and the PN2 setups} the binary semi-major axis and eccentricity remain almost constant with time.

{Since most of the dephasing is accumulated at larger separations, where the coupling with the disk is stronger, we align the relevant pairs of simulations as much as possible, given the temporal resolution limitation in our simulations, in phase (set to $\pi$) at the final time. The ideal approach would be to align the phases exactly at merger but evolving our systems further would significantly increase the computational cost without adding significant information on the dephasing accumulated at larger separations. We then} measure $\dorb$ by comparing the orbital phases of the appropriate pairs of simulations {at the initial time}. We measure $\decc$ by comparing the argument of periapsis $\omega=\tan^{-1}(e_y/e_x)$ of the appropriate pairs of simulations, where $e_x$ and $e_y$ are the $x$ and $y$ components of the binary eccentricity vector $\vec e$. Lastly, under the quasi-circular approximation, we can infer the GW dephasing ($\dGW$) as twice the orbital dephasing, i.e., $\dGW\approx2\dorb^{\rm(GW)}$. {We should in principle compute the dephasing in each eccentric GW harmonic, since our binary is eccentric to begin with. However, the scope here is to report a simple estimate of GW dephasing for intuition and the majority of the power would be anyway contained in the fundamental harmonic (i.e at twice the orbital frequency) for our choice of parameters \citep{Peters1963}.} 

In the next section, we show the results together with the interpretation of our simulations.

\section{Results}\label{Sec:Results}

\begin{figure*}
    \centering
    \includegraphics[width=0.9\linewidth]{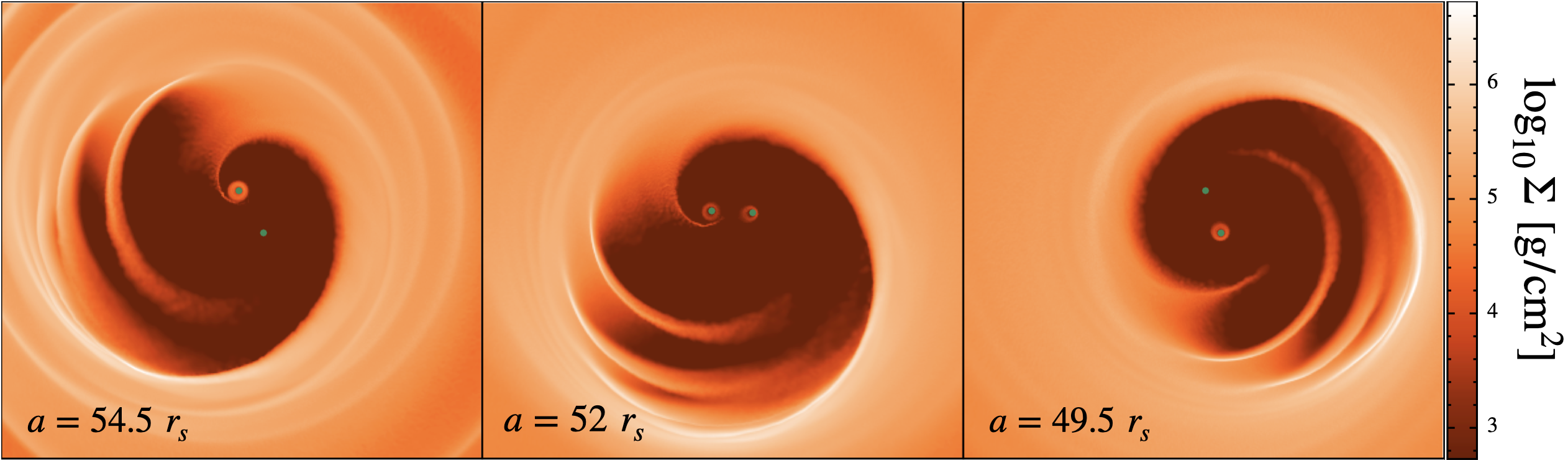}
    \caption{Column density ($\Sigma$) plots at three SMAs: $54.5~r_s$ (left panel), $52~r_s$ (middle panel), and $49.5~r_s$ (right panel) for the binary evolution under both GW and gas. Here $\Sigma$ varies between $\sim10^{3}$-$10^{7}$ g/cm$^{2}$. Both the binary (green dots) and the cavity shrink with time. Moreover, gas inflow inside the cavity creates short-lived mini-disks.}
    \label{fig:disk}
\end{figure*}

We show the disk morphology in terms of column density at three SMAs -- $a=54.5~r_s$, $a=52~r_s$, and $a=49.5~r_s$ -- in Fig.~\ref{fig:disk} for our MR {gas+PN2.5} simulation.\footnote{The disk morphology for {gas+PN2} simulation is similar to the left panel of Fig.~\ref{fig:disk}.} The gas morphology is similar, as expected, to the simulations presented in \cite{Franchini2024} with the addition of short-lived mini-disks owing to our higher resolution. We can clearly see the over-density at the cavity edge, i.e. the lump \citep{Shi2012}, precessing around the binary.

We measure the accretion rate in both our simulations and find a mean Eddington ratio of $\bar{\fedd}\approx1.30$ for {gas+PN2.5} simulation and $\bar{\fedd}\approx1.50$ for the {gas+PN2} case, both consistent with the analytical expectation for {the choice of disk mass ($5\MSun$) we made and assuming} a steady-state {$\alpha-$}disk with the properties we chose \citep{Frank2002}. The slightly higher accretion rate in the {gas+PN2} case is a natural consequence of the fact that the binary does not decouple from the disk because of the lack of GW radiation, therefore the gas can flow inside the cavity and keep feeding the MBHB more effectively.

\begin{figure}
    \centering
    \includegraphics[width=0.45\textwidth]{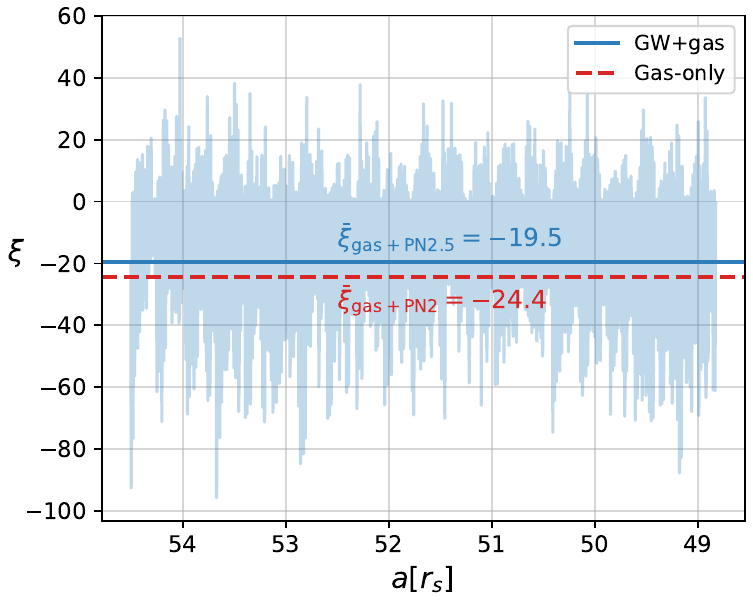}
    \caption{Gas torques onto the binary in terms of $\xi$ as a function of the SMA for {gas+PN2.5} simulation (light blue lines). We show average $\xi$ values for {gas+PN2.5} (solid blue line; $\bar\xi_{\rm {gas+PN2.5}}\approx-19.5$) and {gas+PN2} (dashed red line; $\bar\xi_{\rm gas+PN2}\approx-24.5$) cases, respectively. Note that we have not done any smoothing in plotting $\xi$ for the {gas+PN2.5} run but only interpolation between snapshots.}
    \label{fig:Gastorque}
\end{figure}

We compute both gravitational and accretion torques directly from the simulations. We find that the magnitude of the accretion torque is $T_{\rm acc}\sim10^{-2}T_{\rm grav}$. We show the evolution of the parameter $\xi$ in Fig.~\ref{fig:Gastorque} for {gas+PN2.5} simulations as a function of SMA together with the average values of $\xi$ for both {gas+PN2.5} ($\bar\xi_{\rm {gas+PN2.5}}=-19.5$) and {gas+PN2} ($\bar\xi_{\rm {gas+PN2}}=-24.4$) cases. {Negative torques are consistent with similar aspect ratio HD studies \citep{Tiede2020,Heath2020,Dittmann2022,Franchini2022}}. Unsurprisingly, the mean gas effect is stronger when the binary is only evolving due to the gas torques. Since the value of $\xi$ oscillates around its average value over time, one may approximately infer it as a constant parameter within the range of separations explored in this work.

We also note that the high-frequency or sub-orbital fluctuations in Fig.~\ref{fig:Gastorque} around the mean value may be measurable on their own via GWs, as suggested by analytical studies \citep{Zwick2022,Zwick2024,Zwick2025}. However, recently, \citet{Copparoni2025} demonstrated, using realistic LISA data analysis, that these moderate fluctuations we find are not measurable in GWs, albeit for a much smaller mass ratio BHB compared to the one explored in our work.

\begin{figure*}
    \centering
    \includegraphics[width=0.9\linewidth]{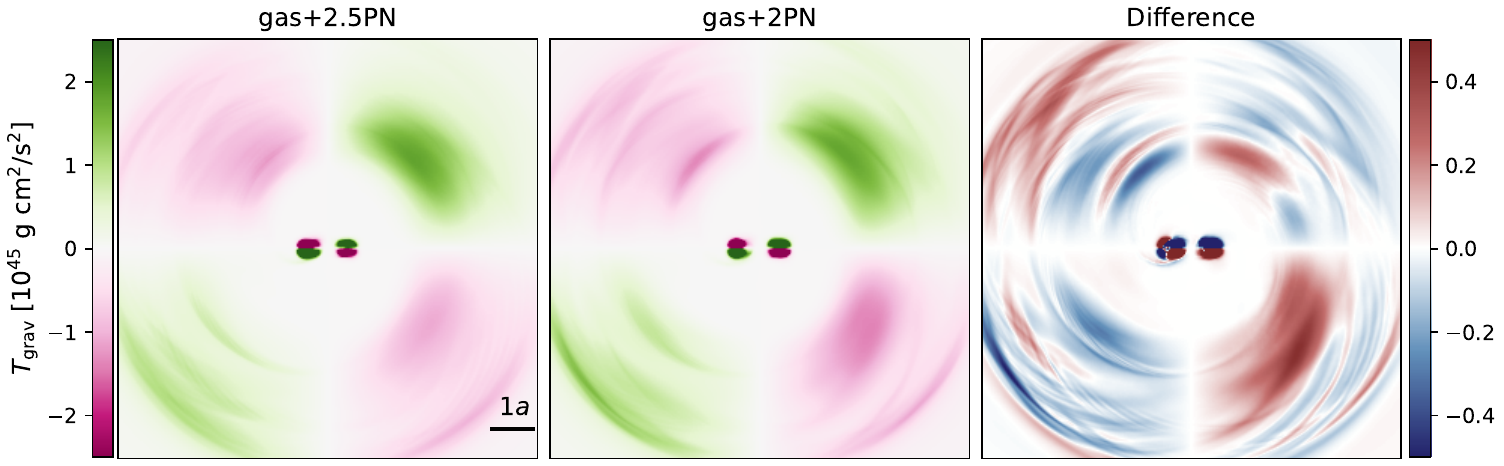}
    \caption{2D projected gravitational torque ($T_{\rm grav}$) distribution between $-5a$ to $5a$ in both axes averaged over $100$ snapshots between $100^{\rm th}$ and $110^{\rm th}$ orbits. We show results from the {gas+PN2.5} run (left panel), the {gas+PN2} run (middle panel), and their difference (right panel). The third panel clearly shows that the {gas+PN2.5} run has slightly more positive torque than the {gas+PN2} simulation.}
    \label{fig:Tmaps}
\end{figure*}

We show torque density maps averages over ten snapshots for the {gas+PN2.5} (left panel) and {gas+PN2} (middle panel) simulations together with their difference (right panel) in Fig.~\ref{fig:Tmaps}. Similar to previous HD studies \citep[see, e.g.][]{Tiede2020,Franchini2023}, we rotate gas particles about the binary's center-of-mass such that the axis connecting the two BHs aligns in the two simulations to do a meaningful comparison between their gas distribution. We chose to average over ten orbits, as it is small enough that the binary is not significantly shrinking and large enough to remove instantaneous features. The first two panels show the contribution from both CBD and mini-disks to the gravitational torque. {The gas in front of the BH tends to drag it (pink color) along while the gas behind it decelerates it (green color).}
The right panel clearly shows that the gravitational torque is larger in the {gas+PN2} simulation as there are more regions where the torque is negative. This is again consistent with the estimate of the $\xi$ parameter in Fig. \ref{fig:Gastorque}.

\subsection{Gas-induced dephasings and LISA observability}
\label{Sec:dephasing}

In this section, we consider different dephasings introduced at the end of \S~\ref{Sec:Postprocess}. We enumerate source-frame gas-induced dephasings that corresponds to elapsed physical time between $54.5~r_s$ and $48.9~r_s$ for the {gas+PN2.5} simulation over 278 orbits or $\sim 1745.8$ rad orbital phase in Table~\ref{table:Dephasings}. We have compared our {PN2.5} simulation with a numerical integration of the full (i.e. up to 2.5 order) PN evolution of the binary using an 8th order Runge-Kutta integrator finding mean relative error of $\lesssim 0.1\%$ within the separations range we explored. This illustrates that our orbital parameters are well measured in different simulations.

\begin{table}
\centering
\renewcommand{\arraystretch}{1.5}
    \begin{tabular}{|C|C|}
        \hline
        {\rm \bf Dephasing}&{\bf Value [rad]}\\
        \hline
        \hline
        \dorb^{\rm(GW)} & -0.007\\
        \hline
        \decc^{\rm(GW)} & -0.012\\
        \hline
        \dorb^{\rm(NoGW)} & -0.014\\
        \hline 
        \decc^{\rm(NoGW)} &-0.010\\
        \hline 
        \dGW & -0.014\\
        \hline 
    \end{tabular}
\caption{Dephasings measured from our simulations: gas-induced orbital $\dorb^{\rm(GW)}$ and precession $\decc^{\rm(GW)}$ dephasings between {gas+PN2.5} vs {PN2.5} simulations and same dephasings but between {gas+PN2} and {PN2} runs ($\dorb^{\rm(NoGW)},\decc^{\rm(NoGW)}$). We also infer gas-induced GW dephasing by doubling $\dorb^{\rm(GW)}$.} 
\label{table:Dephasings}
\renewcommand{\arraystretch}{1}
\end{table}

The comparison between the orbital phase evolution in the {gas+PN2.5} and {PN2.5} simulations, $\dorb^{\rm (GW)}$, allows us to quantify the overall contribution of the different perturbations induced by the gas disk in the GW-driven binary evolution, i.e. its effect on $\dot a $, $\dot e $, and the precession rate.
We can directly measure the dephasing due to the different precession rates $\decc^{\rm (GW)}$ directly from the simulations. However this is  still not completely independent from other changes in the orbital parameters. Moreover, GWs radiation further complicates the picture as it changes the gas morphology, which can further contribute to $\dorb^{\rm (GW)}$ in a non-linear manner.

Comparing the {gas+PN2} and the {PN2} runs can alleviate some of these issues as these runs both neglect the GW-driven fast inspiral, allowing us to measure the dephasing due to  disk-induced binary precession $\decc^{\rm (NoGW)}$ in a system where the semi-major axis and eccentricity of the binary do not significantly evolve with time. 

The results presented in Table \ref{table:Dephasings} show that the orbital dephasing measured in the {gas+PN2.5} simulation (first row) is smaller than the same dephasing inferred in absence of GW emission (third row). This is consistent with the effect of the gas becoming weaker as the binary inspirals. The estimate of dephasing due to the binary precession in the second row is affected by the change in binary eccentricity induced by GW emission and is therefore not directly comparable to the term in the first row. The precession-induced dephasing calculated from the NoGW simulations (fourth row) is more similar to the term in the first row. This seems to indicate that the main contributor to the binary orbital dephasing is the disk-induced precession of the binary eccentricity vector. However we restrain to draw such a strong conclusion as longer simulations together with a larger exploration of the parameter space is needed to understand the role that non-linear effects play in the ultimate orbital dephasing in GWs. Furthermore, we note that all the dephasing inferred from the simulations are subjected to the particles noise and to the precise alignment of the orbital phase and argument of periapsis at the end of the simulations. 

If we compute the gas-induced GW dephasing analytically using only the gas torques by linearly adding SMA rates due to GWs and gas \citep{Garg2022,Dittmann2023,Duffell2024} then we get $\dGW^{(\xi)}=-0.095$ rad, which is a factor of $7$ higher than our direct estimate of $\dGW=-0.014$ rad. This difference could be due to i) the inadequacy of the analytical prescription, ii) the approximation made by simply linearly adding two terms (i.e. GW-driven  and gas-driven SMA change) that mutually affect each other and iii) different numerical approaches. Follow-up simulations might shed more light onto the nature of this discrepancy. We note that our estimate of precession-induced dephasing of $\decc^{\rm (GW)}=-0.012$ rad over 278 orbits is in magnitude comparable to the analytical prescription by \citet{Tiede2024a} that yields dephasing of $\sim0.010$ rad. This indicates that the choice of numerical method has a limited effect on the precession-induced dephasing.

Lastly, if we consider our fiducial MBHB at $z=1$ then LISA should observe this event with SNR$\sim1300$ \citep{Garg2024a}. {As a first-order test,} a given absolute dephasing needs to be higher than $\sim8/$SNR $\approx0.006$ rad \citep{Kocsis2011,Derdzinski2021,Garg2022} to be detectable{, where $8$ is a typical GW event SNR threshold to be observable \citep{Bonetti2019}}. Therefore, our measured gas-induced dephasing of $-0.014$ rad should be observable. {While a better check will be do to Bayesian inference \citep{Garg2024b}, it is beyond the scope of this work.}

\section{Discussion and Conclusion}

We studied the interaction of an equal mass MBHB with its surrounding geometrically thin ($H/R=0.03$) CBD during the late in-spiral stage, with and without concurrent GW emission, using 3D hydrodynamical simulations with a live binary. This approach provided us with the first direct measurement of how surrounding gas torques the binary when its in-spiral is already governed by GW emission, by means of the estimate of the $\xi$ parameter in Eq.~\eqref{eq:xi}. 
We find that $\xi$ is $\mathcal{O}(10)$ stronger than in some of the previously explored scale-free/sub-pc regime. However, we caution that the comparison may not be fair as previous simulations carried out for the larger separation regime are predominantly 2D, assume a fixed binary orbit modeled under Newtonian dynamics, explore larger values of the viscosity parameter $\nu$ \citep{Dittmann2022}, and compute the effect of the energy and angular momentum loss by GW radiation in post-processing \citep{Tang2018}. 

Note that if we compared our value of $\xi$ in the {gas+PN2} simulation with the results presented in \cite{Tiede2025} with the same value of viscosity that we have and the highest resolution they explored, we find the difference to be a factor of two. We also find that the measured time-averaged torque $\xi$ becomes weaker with higher resolutions (see Fig.~\ref{fig:Torque_res}), which is the opposite trend to the one found by \citet{Tiede2025}. The different behavior might be due to our 3D live-orbit PN treatment with respect to their 2D fixed-orbit Newtonian simulations and to the different Mach number, i.e. $\mathcal{M}\approx33$, employed in our simulations rather than their $\mathcal{M}=40$.
In particular, \citet{Duffell2024} showed that 3D calculations give different  magnitude torques compared to 2D and \cite{Franchini2023} argued that fixing the binary orbit leads again to a different gravitational torque. 
Moreover, we find that, consistently with our expectations, $\xi$ is slightly weaker for {gas+PN2.5} simulations ($\bar \xi_{\rm {gas+PN2.5}}\sim-19.5$) with respect to {gas+PN2} study ($\bar \xi_{\rm {gas+PN2}}\sim-24.4$), as shown in Fig.~\ref{fig:Gastorque}. This is expected since the binary is decoupling from the gas in the {gas+PN2.5} simulation and therefore the effect of gas weakens with time.

It is to be noted that we employ PN corrections as an approximation since we do not run general relativistic magnetohydrodynamical (GRMHD) simulations. Current GRMHD simulations only study the binary evolution just a few days before merger \citep{Gutierrez2022,Avara2024} to integrate only a few orbits because of the prohibitive computational cost. Therefore, since the majority of the gas-driven effects on the binary in-spiral occurs at separations $a\gtrsim48.9~r_s$, our approach is currently the best available method to investigate the orbital dephasing due to the presence of the gas and where PN corrections may be adequate.

A caveat in this work is that we assume that the gas morphology will look the same at our initial separation for all our {gas+PN2.5} and {gas+PN2} simulations. In principle, one would need to start the binary at sufficiently large separations, approaching parsecs, that the GW radiation is completely negligible, and quantify the difference in the morphology of the gas distribution while the binary shrinks as the GW radiation gradually ensues relative to a case in which it is neglected. This is currently not possible due to the prohibitive computational costs. If anything, by starting the binary with the same initial condition in the {gas+PN2.5} and {gas+PN2} simulation at a separation at which GW radiation is already taking place, we are erring on the side of underestimating the back-reaction of the gas to the GW emission, which translates into a conservative estimate of the cross-term.

Another possible caveat in our work is that our binary is moderately eccentric ($\sim0.3$) just a few years before the merger, which will require the eccentricity to be extremely high when GWs take over at milli-pc scales. However, since $e\sim0.3$ arises naturally from our initial condition requiring a steady-state disk  before setting the physical scale of $a\sim55r_s$, the only truly realistic way to initialize the system is to evolve the binary starting from a much wider separation. This, however, would increase the computational cost dramatically. We plan to investigate alternative procedures in the setup of the simulations in order to reduce the computational cost and mitigate this issue in the future. 

In summary, our results can facilitate the modeling of gas effects perturbing GW waveforms, which in turn will allow to better quantify how effectively LISA can place constraints on the environment of MBHBs, eventually opening the pathway for more informed synergies between GWs and EM observations. Furthermore, our work, being the first of its kind with PN dynamics and a live binary in 3D, while still assuming a simple isothermal equation of state, provides a starting point for future hydrodynamical studies with additional physics, including, for example, more realistic thermodynamics.

\section*{Acknowledgments}
MG acknowledge support from the Swiss National Science Foundation (SNSF) under the grant 200020\_192092 and ``GW-learn" grant agreement CRSII5 213497. AF acknowledges support provided by the ``GW-learn" grant agreement CRSII5 213497 and the Tomalla Foundation. AL acknowledges support by PRIN MUR “2022935STW". We thank the anonymous referee for helpful comments that improved this work. We futher thank Rohit Chandramouli, Andrea Derdzinski, Alexander Dittmann, Callum Fairbairn, Zoltan Haiman, Laura Sberna, Connar Rowan, Christopher Tiede, and Lorenz Zwick for useful discussions. The authors also acknowledge use of the NumPy \citep{harris2020array} and Matplotlib \citep{Hunter2007}.

\bibliography{TorqueLISA}

\begin{thebibliography}{}
\expandafter\ifx\csname natexlab\endcsname\relax\def\natexlab#1{#1}\fi
\providecommand{\url}[1]{\href{#1}{#1}}
\providecommand{\dodoi}[1]{doi:~\href{http://doi.org/#1}{\nolinkurl{#1}}}
\providecommand{\doeprint}[1]{\href{http://ascl.net/#1}{\nolinkurl{http://ascl.net/#1}}}
\providecommand{\doarXiv}[1]{\href{https://arxiv.org/abs/#1}{\nolinkurl{https://arxiv.org/abs/#1}}}

\bibitem[{{Amaro-Seoane} {et~al.}(2017){Amaro-Seoane}, {Audley}, {Babak}, {Baker}, {Barausse}, {Bender}, {Berti}, {Binetruy}, {Born}, {Bortoluzzi}, {Camp}, {Caprini}, {Cardoso}, {Colpi}, {Conklin}, {Cornish}, {Cutler}, {Danzmann}, {Dolesi}, {Ferraioli}, {Ferroni}, {Fitzsimons}, {Gair}, {Gesa Bote}, {Giardini}, {Gibert}, {Grimani}, {Halloin}, {Heinzel}, {Hertog}, {Hewitson}, {Holley-Bockelmann}, {Hollington}, {Hueller}, {Inchauspe}, {Jetzer}, {Karnesis}, {Killow}, {Klein}, {Klipstein}, {Korsakova}, {Larson}, {Livas}, {Lloro}, {Man}, {Mance}, {Martino}, {Mateos}, {McKenzie}, {McWilliams}, {Miller}, {Mueller}, {Nardini}, {Nelemans}, {Nofrarias}, {Petiteau}, {Pivato}, {Plagnol}, {Porter}, {Reiche}, {Robertson}, {Robertson}, {Rossi}, {Russano}, {Schutz}, {Sesana}, {Shoemaker}, {Slutsky}, {Sopuerta}, {Sumner}, {Tamanini}, {Thorpe}, {Troebs}, {Vallisneri}, {Vecchio}, {Vetrugno}, {Vitale}, {Volonteri}, {Wanner}, {Ward}, {Wass}, {Weber}, {Ziemer}, \& {Zweifel}}]{AmaroSeoane2017}
{Amaro-Seoane}, P., {Audley}, H., {Babak}, S., {et~al.} 2017, arXiv e-prints, arXiv:1702.00786.
\newblock \doarXiv{1702.00786}

\bibitem[{{Amaro-Seoane} {et~al.}(2023){Amaro-Seoane}, {Andrews}, {Arca Sedda}, {Askar}, {Baghi}, {Balasov}, {Bartos}, {Bavera}, {Bellovary}, {Berry}, {Berti}, {Bianchi}, {Blecha}, {Blondin}, {Bogdanovi{\'c}}, {Boissier}, {Bonetti}, {Bonoli}, {Bortolas}, {Breivik}, {Capelo}, {Caramete}, {Cattorini}, {Charisi}, {Chaty}, {Chen}, {Chru{\'s}li{\'n}ska}, {Chua}, {Church}, {Colpi}, {D'Orazio}, {Danielski}, {Davies}, {Dayal}, {De Rosa}, {Derdzinski}, {Destounis}, {Dotti}, {Dutan}, {Dvorkin}, {Fabj}, {Foglizzo}, {Ford}, {Fouvry}, {Franchini}, {Fragos}, {Fryer}, {Gaspari}, {Gerosa}, {Graziani}, {Groot}, {Habouzit}, {Haggard}, {Haiman}, {Han}, {Istrate}, {Johansson}, {Khan}, {Kimpson}, {Kokkotas}, {Kong}, {Korol}, {Kremer}, {Kupfer}, {Lamberts}, {Larson}, {Lau}, {Liu}, {Lloyd-Ronning}, {Lodato}, {Lupi}, {Ma}, {Maccarone}, {Mandel}, {Mangiagli}, {Mapelli}, {Mathis}, {Mayer}, {McGee}, {McKernan}, {Miller}, {Mota}, {Mumpower}, {Nasim}, {Nelemans}, {Noble}, {Pacucci}, {Panessa}, {Paschalidis}, {Pfister}, {Porquet},
  {Quenby}, {Ricarte}, {R{\"o}pke}, {Regan}, {Rosswog}, {Ruiter}, {Ruiz}, {Runnoe}, {Schneider}, {Schnittman}, {Secunda}, {Sesana}, {Seto}, {Shao}, {Shapiro}, {Sopuerta}, {Stone}, {Suvorov}, {Tamanini}, {Tamfal}, {Tauris}, {Temmink}, {Tomsick}, {Toonen}, {Torres-Orjuela}, {Toscani}, {Tsokaros}, {Unal}, {V{\'a}zquez-Aceves}, {Valiante}, {van Putten}, {van Roestel}, {Vignali}, {Volonteri}, {Wu}, {Younsi}, {Yu}, {Zane}, {Zwick}, {Antonini}, {Baibhav}, {Barausse}, {Bonilla Rivera}, {Branchesi}, {Branduardi-Raymont}, {Burdge}, {Chakraborty}, {Cuadra}, {Dage}, {Davis}, {de Mink}, {Decarli}, {Doneva}, {Escoffier}, {Gandhi}, {Haardt}, {Lousto}, {Nissanke}, {Nordhaus}, {O'Shaughnessy}, {Portegies Zwart}, {Pound}, {Schussler}, {Sergijenko}, {Spallicci}, {Vernieri}, \& {Vigna-G{\'o}mez}}]{AmaroSeoane2023}
{Amaro-Seoane}, P., {Andrews}, J., {Arca Sedda}, M., {et~al.} 2023, Living Reviews in Relativity, 26, 2, \dodoi{10.1007/s41114-022-00041-y}

\bibitem[{{Armitage} \& {Natarajan}(2002)}]{ArmitageNatarajan2002}
{Armitage}, P.~J., \& {Natarajan}, P. 2002, \apjl, 567, L9, \dodoi{10.1086/339770}

\bibitem[{{Artymowicz} \& {Lubow}(1994)}]{artymowicz1994}
{Artymowicz}, P., \& {Lubow}, S.~H. 1994, \apj, 421, 651, \dodoi{10.1086/173679}

\bibitem[{{Artymowicz} \& {Lubow}(1996)}]{artymowicz1996}
---. 1996, \apjl, 467, L77, \dodoi{10.1086/310200}

\bibitem[{{Avara} {et~al.}(2024){Avara}, {Krolik}, {Campanelli}, {Noble}, {Bowen}, \& {Ryu}}]{Avara2024}
{Avara}, M.~J., {Krolik}, J.~H., {Campanelli}, M., {et~al.} 2024, \apj, 974, 242, \dodoi{10.3847/1538-4357/ad5bda}

\bibitem[{{Bate} {et~al.}(1995){Bate}, {Bonnell}, \& {Price}}]{Bate1995}
{Bate}, M.~R., {Bonnell}, I.~A., \& {Price}, N.~M. 1995, \mnras, 277, 362, \dodoi{10.1093/mnras/277.2.362}

\bibitem[{{Begelman} {et~al.}(1980){Begelman}, {Blandford}, \& {Rees}}]{Begelman1980}
{Begelman}, M.~C., {Blandford}, R.~D., \& {Rees}, M.~J. 1980, \nat, 287, 307, \dodoi{10.1038/287307a0}

\bibitem[{{Blaes} {et~al.}(2002){Blaes}, {Lee}, \& {Socrates}}]{Blaes2002}
{Blaes}, O., {Lee}, M.~H., \& {Socrates}, A. 2002, \apj, 578, 775, \dodoi{10.1086/342655}

\bibitem[{{Blanchet}(2014)}]{Blanchet2014}
{Blanchet}, L. 2014, Living Reviews in Relativity, 17, 2, \dodoi{10.12942/lrr-2014-2}

\bibitem[{{Bonetti} {et~al.}(2019){Bonetti}, {Sesana}, {Haardt}, {Barausse}, \& {Colpi}}]{Bonetti2019}
{Bonetti}, M., {Sesana}, A., {Haardt}, F., {Barausse}, E., \& {Colpi}, M. 2019, \mnras, 486, 4044, \dodoi{10.1093/mnras/stz903}

\bibitem[{{Colpi} {et~al.}(2024){Colpi}, {Danzmann}, {Hewitson}, {Holley-Bockelmann}, {Jetzer}, {Nelemans}, {Petiteau}, {Shoemaker}, {Sopuerta}, {Stebbins}, {Tanvir}, {Ward}, {Weber}, {Thorpe}, {Daurskikh}, {Deep}, {Fern{\'a}ndez N{\'u}{\~n}ez}, {Garc{\'\i}a Marirrodriga}, {Gehler}, {Halain}, {Jennrich}, {Lammers}, {Larra{\~n}aga}, {Lieser}, {L{\"u}tzgendorf}, {Martens}, {Mondin}, {Piris Ni{\~n}o}, {Amaro-Seoane}, {Arca Sedda}, {Auclair}, {Babak}, {Baghi}, {Baibhav}, {Baker}, {Bayle}, {Berry}, {Berti}, {Boileau}, {Bonetti}, {Brito}, {Buscicchio}, {Calcagni}, {Capelo}, {Caprini}, {Caputo}, {Castelli}, {Chen}, {Chen}, {Chua}, {Davies}, {Derdzinski}, {Domcke}, {Doneva}, {Dvorkin}, {Mar{\'\i}a Ezquiaga}, {Gair}, {Haiman}, {Harry}, {Hartwig}, {Hees}, {Heffernan}, {Husa}, {Izquierdo-Villalba}, {Karnesis}, {Klein}, {Korol}, {Korsakova}, {Kupfer}, {Laghi}, {Lamberts}, {Larson}, {Le Jeune}, {Lewicki}, {Littenberg}, {Madge}, {Mangiagli}, {Marsat}, {Vilchez}, {Maselli}, {Mathews}, {van de Meent}, {Muratore}, {Nardini},
  {Pani}, {Peloso}, {Pieroni}, {Pound}, {Quelquejay-Leclere}, {Ricciardone}, {Rossi}, {Sartirana}, {Savalle}, {Sberna}, {Sesana}, {Shoemaker}, {Slutsky}, {Sotiriou}, {Speri}, {Staab}, {Steer}, {Tamanini}, {Tasinato}, {Torrado}, {Torres-Orjuela}, {Toubiana}, {Vallisneri}, {Vecchio}, {Volonteri}, {Yagi}, \& {Zwick}}]{Colpi2024}
{Colpi}, M., {Danzmann}, K., {Hewitson}, M., {et~al.} 2024, arXiv e-prints, arXiv:2402.07571, \dodoi{10.48550/arXiv.2402.07571}

\bibitem[{{Copparoni} {et~al.}(2025){Copparoni}, {Barausse}, {Speri}, {Sberna}, \& {Derdzinski}}]{Copparoni2025}
{Copparoni}, L., {Barausse}, E., {Speri}, L., {Sberna}, L., \& {Derdzinski}, A. 2025, \prd, 111, 104079, \dodoi{10.1103/PhysRevD.111.104079}

\bibitem[{{Cuadra} {et~al.}(2009){Cuadra}, {Armitage}, {Alexander}, \& {Begelman}}]{Cuadra2009}
{Cuadra}, J., {Armitage}, P.~J., {Alexander}, R.~D., \& {Begelman}, M.~C. 2009, \mnras, 393, 1423, \dodoi{10.1111/j.1365-2966.2008.14147.x}

\bibitem[{{Derdzinski} {et~al.}(2021){Derdzinski}, {D'Orazio}, {Duffell}, {Haiman}, \& {MacFadyen}}]{Derdzinski2021}
{Derdzinski}, A., {D'Orazio}, D., {Duffell}, P., {Haiman}, Z., \& {MacFadyen}, A. 2021, \mnras, 501, 3540, \dodoi{10.1093/mnras/staa3976}

\bibitem[{{Dittmann} \& {Ryan}(2022)}]{Dittmann2022}
{Dittmann}, A.~J., \& {Ryan}, G. 2022, \mnras, 513, 6158, \dodoi{10.1093/mnras/stac935}

\bibitem[{{Dittmann} {et~al.}(2023){Dittmann}, {Ryan}, \& {Miller}}]{Dittmann2023}
{Dittmann}, A.~J., {Ryan}, G., \& {Miller}, M.~C. 2023, \apjl, 949, L30, \dodoi{10.3847/2041-8213/acd183}

\bibitem[{D'Orazio {et~al.}(2016)D'Orazio, Haiman, Duffell, MacFadyen, \& Farris}]{DOrazio2016}
D'Orazio, D.~J., Haiman, Z., Duffell, P., MacFadyen, A.~I., \& Farris, B.~D. 2016, Mon. Not. Roy. Astron. Soc., 459, 2379, \dodoi{10.1093/mnras/stw792}

\bibitem[{{Duffell} {et~al.}(2024){Duffell}, {Dittmann}, {D'Orazio}, {Franchini}, {Kratter}, {Penzlin}, {Ragusa}, {Siwek}, {Tiede}, {Wang}, {Zrake}, {Dempsey}, {Haiman}, {Lupi}, {Pirog}, \& {Ryan}}]{Duffell2024}
{Duffell}, P.~C., {Dittmann}, A.~J., {D'Orazio}, D.~J., {et~al.} 2024, \apj, 970, 156, \dodoi{10.3847/1538-4357/ad5a7e}

\bibitem[{{Escala} {et~al.}(2004){Escala}, {Larson}, {Coppi}, \& {Mardones}}]{Escala2004}
{Escala}, A., {Larson}, R.~B., {Coppi}, P.~S., \& {Mardones}, D. 2004, \apj, 607, 765, \dodoi{10.1086/386278}

\bibitem[{{Farris} {et~al.}(2014){Farris}, {Duffell}, {MacFadyen}, \& {Haiman}}]{Farris2014}
{Farris}, B.~D., {Duffell}, P., {MacFadyen}, A.~I., \& {Haiman}, Z. 2014, \apj, 783, 134, \dodoi{10.1088/0004-637X/783/2/134}

\bibitem[{{Franchini} {et~al.}(2024){Franchini}, {Bonetti}, {Lupi}, \& {Sesana}}]{Franchini2024}
{Franchini}, A., {Bonetti}, M., {Lupi}, A., \& {Sesana}, A. 2024, \aap, 686, A288, \dodoi{10.1051/0004-6361/202449206}

\bibitem[{{Franchini} {et~al.}(2022){Franchini}, {Lupi}, \& {Sesana}}]{Franchini2022}
{Franchini}, A., {Lupi}, A., \& {Sesana}, A. 2022, \apjl, 929, L13, \dodoi{10.3847/2041-8213/ac63a2}

\bibitem[{{Franchini} {et~al.}(2023){Franchini}, {Lupi}, {Sesana}, \& {Haiman}}]{Franchini2023}
{Franchini}, A., {Lupi}, A., {Sesana}, A., \& {Haiman}, Z. 2023, \mnras, 522, 1569, \dodoi{10.1093/mnras/stad1070}

\bibitem[{{Franchini} {et~al.}(2021){Franchini}, {Sesana}, \& {Dotti}}]{Franchini2021}
{Franchini}, A., {Sesana}, A., \& {Dotti}, M. 2021, \mnras, 507, 1458, \dodoi{10.1093/mnras/stab2234}

\bibitem[{{Frank} {et~al.}(2002){Frank}, {King}, \& {Raine}}]{Frank2002}
{Frank}, J., {King}, A., \& {Raine}, D.~J. 2002, {Accretion Power in Astrophysics: Third Edition}

\bibitem[{{Garg} {et~al.}(2024{\natexlab{a}}){Garg}, {Derdzinski}, {Tiwari}, {Gair}, \& {Mayer}}]{Garg2024b}
{Garg}, M., {Derdzinski}, A., {Tiwari}, S., {Gair}, J., \& {Mayer}, L. 2024{\natexlab{a}}, \mnras, 532, 4060, \dodoi{10.1093/mnras/stae1764}

\bibitem[{{Garg} {et~al.}(2022){Garg}, {Derdzinski}, {Zwick}, {Capelo}, \& {Mayer}}]{Garg2022}
{Garg}, M., {Derdzinski}, A., {Zwick}, L., {Capelo}, P.~R., \& {Mayer}, L. 2022, \mnras, 517, 1339, \dodoi{10.1093/mnras/stac2711}

\bibitem[{{Garg} {et~al.}(2024{\natexlab{b}}){Garg}, {Sberna}, {Speri}, {Duque}, \& {Gair}}]{Garg2024d}
{Garg}, M., {Sberna}, L., {Speri}, L., {Duque}, F., \& {Gair}, J. 2024{\natexlab{b}}, \mnras, 535, 3283, \dodoi{10.1093/mnras/stae2605}

\bibitem[{{Garg} {et~al.}(2024{\natexlab{c}}){Garg}, {Tiede}, \& {D'Orazio}}]{Garg2024c}
{Garg}, M., {Tiede}, C., \& {D'Orazio}, D.~J. 2024{\natexlab{c}}, \mnras, 534, 3705, \dodoi{10.1093/mnras/stae2357}

\bibitem[{{Garg} {et~al.}(2024{\natexlab{d}}){Garg}, {Tiwari}, {Derdzinski}, {Baker}, {Marsat}, \& {Mayer}}]{Garg2024a}
{Garg}, M., {Tiwari}, S., {Derdzinski}, A., {et~al.} 2024{\natexlab{d}}, \mnras, 528, 4176, \dodoi{10.1093/mnras/stad3477}

\bibitem[{{Gong} {et~al.}(2021){Gong}, {Xu}, {Gui}, {Huang}, \& {Lau}}]{Gong2021}
{Gong}, X., {Xu}, S., {Gui}, S., {Huang}, S., \& {Lau}, Y.-K. 2021, in Handbook of Gravitational Wave Astronomy (Springer Singapore), 24, \dodoi{10.1007/978-981-15-4702-7_24-1}

\bibitem[{{Guti{\'e}rrez} {et~al.}(2022){Guti{\'e}rrez}, {Combi}, {Noble}, {Campanelli}, {Krolik}, {L{\'o}pez Armengol}, \& {Garc{\'\i}a}}]{Gutierrez2022}
{Guti{\'e}rrez}, E.~M., {Combi}, L., {Noble}, S.~C., {et~al.} 2022, \apj, 928, 137, \dodoi{10.3847/1538-4357/ac56de}

\bibitem[{{Haiman} {et~al.}(2009){Haiman}, {Kocsis}, \& {Menou}}]{Haiman2009}
{Haiman}, Z., {Kocsis}, B., \& {Menou}, K. 2009, \apj, 700, 1952, \dodoi{10.1088/0004-637X/700/2/1952}

\bibitem[{Harris {et~al.}(2020)Harris, Millman, van~der Walt, Gommers, Virtanen, Cournapeau, Wieser, Taylor, Berg, Smith, Kern, Picus, Hoyer, van Kerkwijk, Brett, Haldane, del R{\'{i}}o, Wiebe, Peterson, G{\'{e}}rard-Marchant, Sheppard, Reddy, Weckesser, Abbasi, Gohlke, \& Oliphant}]{harris2020array}
Harris, C.~R., Millman, K.~J., van~der Walt, S.~J., {et~al.} 2020, Nature, 585, 357, \dodoi{10.1038/s41586-020-2649-2}

\bibitem[{{Heath} \& {Nixon}(2020)}]{Heath2020}
{Heath}, R.~M., \& {Nixon}, C.~J. 2020, \aap, 641, A64, \dodoi{10.1051/0004-6361/202038548}

\bibitem[{{Hoffman} \& {Loeb}(2007)}]{Hoffman2007}
{Hoffman}, L., \& {Loeb}, A. 2007, \mnras, 377, 957, \dodoi{10.1111/j.1365-2966.2007.11694.x}

\bibitem[{{Hopkins}(2015)}]{Hopkins2015}
{Hopkins}, P.~F. 2015, \mnras, 450, 53, \dodoi{10.1093/mnras/stv195}

\bibitem[{Hunter(2007)}]{Hunter2007}
Hunter, J.~D. 2007, Computing in Science \& Engineering, 9, 90, \dodoi{10.1109/MCSE.2007.55}

\bibitem[{{Khan} {et~al.}(2018){Khan}, {Capelo}, {Mayer}, \& {Berczik}}]{Khan2018}
{Khan}, F.~M., {Capelo}, P.~R., {Mayer}, L., \& {Berczik}, P. 2018, \apj, 868, 97, \dodoi{10.3847/1538-4357/aae77b}

\bibitem[{{Khan} {et~al.}(2011){Khan}, {Just}, \& {Merritt}}]{Khan2011}
{Khan}, F.~M., {Just}, A., \& {Merritt}, D. 2011, \apj, 732, 89, \dodoi{10.1088/0004-637X/732/2/89}

\bibitem[{{Kocsis} {et~al.}(2011){Kocsis}, {Yunes}, \& {Loeb}}]{Kocsis2011}
{Kocsis}, B., {Yunes}, N., \& {Loeb}, A. 2011, \prd, 84, 024032, \dodoi{10.1103/PhysRevD.84.024032}

\bibitem[{{Li} {et~al.}(2025){Li}, {Liu}, {Torres-Orjuela}, {Chen}, {Inayoshi}, {Wang}, {Hu}, {Amaro-Seoane}, {Askar}, {Bambi}, {Capelo}, {Chen}, {Chua}, {Cond{\'e}s-Bre{\~n}a}, {Dai}, {Das}, {Derdzinski}, {Fan}, {Fujii}, {Gao}, {Garg}, {Ge}, {Giersz}, {Huang}, {Hypki}, {Liang}, {Liu}, {Liu}, {Liu}, {Liu}, {Mayer}, {Napolitano}, {Peng}, {Shao}, {Shashank}, {Shen}, {Tagawa}, {Tanikawa}, {Toscani}, {V{\'a}zquez-Aceves}, {Wang}, {Wang}, {Yi}, {Zhang}, {Zhang}, {Zhu}, {Zwick}, {Huang}, {Mei}, {Wang}, {Xie}, {Zhang}, \& {Luo}}]{Li2025}
{Li}, E.-K., {Liu}, S., {Torres-Orjuela}, A., {et~al.} 2025, Reports on Progress in Physics, 88, 056901, \dodoi{10.1088/1361-6633/adc9be}

\bibitem[{{Liptai} \& {Price}(2019)}]{Liptai2019}
{Liptai}, D., \& {Price}, D.~J. 2019, \mnras, 485, 819, \dodoi{10.1093/mnras/stz111}

\bibitem[{Mangiagli {et~al.}(2022)Mangiagli, Caprini, Volonteri, Marsat, Vergani, Tamanini, \& Inchausp\'e}]{Mangiagli2022}
Mangiagli, A., Caprini, C., Volonteri, M., {et~al.} 2022, Phys. Rev. D, 106, 103017, \dodoi{10.1103/PhysRevD.106.103017}

\bibitem[{{Mayer}(2013)}]{Mayer2013}
{Mayer}, L. 2013, Classical and Quantum Gravity, 30, 244008, \dodoi{10.1088/0264-9381/30/24/244008}

\bibitem[{{Peters} \& {Mathews}(1963)}]{Peters1963}
{Peters}, P.~C., \& {Mathews}, J. 1963, Physical Review, 131, 435, \dodoi{10.1103/PhysRev.131.435}

\bibitem[{{Preto} {et~al.}(2011){Preto}, {Berentzen}, {Berczik}, \& {Spurzem}}]{Preto2011}
{Preto}, M., {Berentzen}, I., {Berczik}, P., \& {Spurzem}, R. 2011, \apjl, 732, L26, \dodoi{10.1088/2041-8205/732/2/L26}

\bibitem[{{Quinlan}(1996)}]{Quinlan1996}
{Quinlan}, G.~D. 1996, \na, 1, 35, \dodoi{10.1016/S1384-1076(96)00003-6}

\bibitem[{{Shakura} \& {Sunyaev}(1973)}]{ShakuraSunyaev1973}
{Shakura}, N.~I., \& {Sunyaev}, R.~A. 1973, \aap, 500, 33

\bibitem[{{Shi} {et~al.}(2012){Shi}, {Krolik}, {Lubow}, \& {Hawley}}]{Shi2012}
{Shi}, J.-M., {Krolik}, J.~H., {Lubow}, S.~H., \& {Hawley}, J.~F. 2012, \apj, 749, 118, \dodoi{10.1088/0004-637X/749/2/118}

\bibitem[{{Souza Lima} {et~al.}(2020){Souza Lima}, {Mayer}, {Capelo}, {Bortolas}, \& {Quinn}}]{SouzaLima2020}
{Souza Lima}, R., {Mayer}, L., {Capelo}, P.~R., {Bortolas}, E., \& {Quinn}, T.~R. 2020, \apj, 899, 126, \dodoi{10.3847/1538-4357/aba624}

\bibitem[{{Tang} {et~al.}(2018){Tang}, {Haiman}, \& {MacFadyen}}]{Tang2018}
{Tang}, Y., {Haiman}, Z., \& {MacFadyen}, A. 2018, \mnras, 476, 2249, \dodoi{10.1093/mnras/sty423}

\bibitem[{{Tiede} {et~al.}(2024){Tiede}, {D'Orazio}, {Zwick}, \& {Duffell}}]{Tiede2024a}
{Tiede}, C., {D'Orazio}, D.~J., {Zwick}, L., \& {Duffell}, P.~C. 2024, \apj, 964, 46, \dodoi{10.3847/1538-4357/ad2613}

\bibitem[{{Tiede} {et~al.}(2020){Tiede}, {Zrake}, {MacFadyen}, \& {Haiman}}]{Tiede2020}
{Tiede}, C., {Zrake}, J., {MacFadyen}, A., \& {Haiman}, Z. 2020, \apj, 900, 43, \dodoi{10.3847/1538-4357/aba432}

\bibitem[{{Tiede} {et~al.}(2025){Tiede}, {Zrake}, {MacFadyen}, \& {Haiman}}]{Tiede2025}
---. 2025, \apj, 984, 144, \dodoi{10.3847/1538-4357/adc727}

\bibitem[{{Zwick} {et~al.}(2022){Zwick}, {Derdzinski}, {Garg}, {Capelo}, \& {Mayer}}]{Zwick2022}
{Zwick}, L., {Derdzinski}, A., {Garg}, M., {Capelo}, P.~R., \& {Mayer}, L. 2022, \mnras, 511, 6143, \dodoi{10.1093/mnras/stac299}

\bibitem[{{Zwick} {et~al.}(2024){Zwick}, {Tiede}, {Trani}, {Derdzinski}, {Haiman}, {D'Orazio}, \& {Samsing}}]{Zwick2024}
{Zwick}, L., {Tiede}, C., {Trani}, A.~A., {et~al.} 2024, \prd, 110, 103005, \dodoi{10.1103/PhysRevD.110.103005}

\bibitem[{{Zwick} {et~al.}(2025){Zwick}, {Hendriks}, {O'Neill}, {Tak{\'a}tsy}, {Kirkeberg}, {Tiede}, {Stegmann}, {Samsing}, \& {D'Orazio}}]{Zwick2025}
{Zwick}, L., {Hendriks}, K., {O'Neill}, D., {et~al.} 2025, \prd, 112, 063005, \dodoi{10.1103/lz7k-bvjf}

\end{thebibliography}
\bibliographystyle{aasjournal}
\normalsize

\appendix
\setcounter{figure}{0}                       
\renewcommand\thefigure{A\arabic{figure}}   
\section{Resolution study}\label{App:Res}

\begin{figure}[h]
    \centering
    \includegraphics[width=0.45\textwidth]{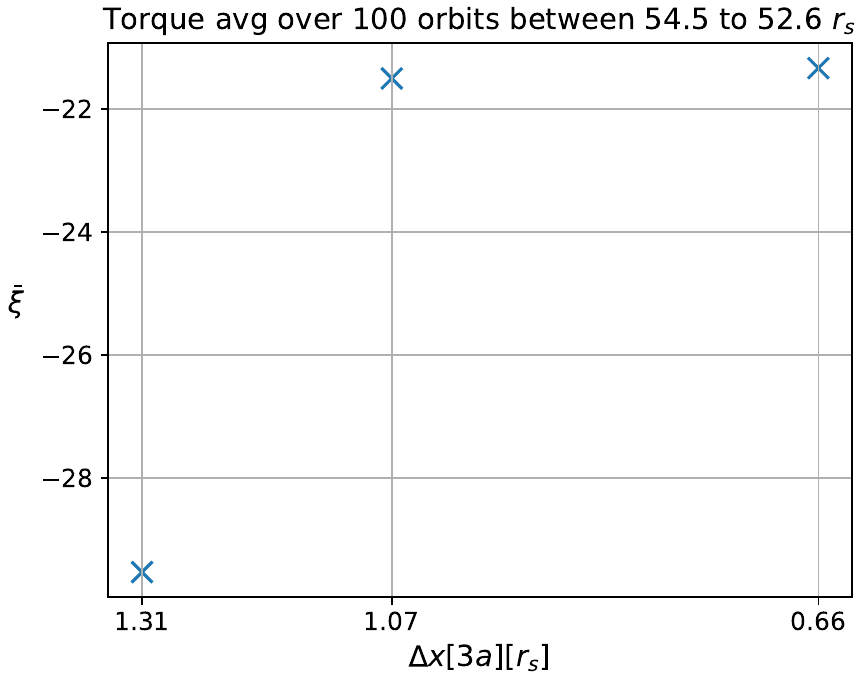}
    \caption{Average torque value (blue cross) expressed in terms of $\bar\xi$ over initial 100 orbits of {gas+PN2.5} simulations for three different resolutions: LR with $\Delta x[3a]={1.31~r_s}$, MR with $\Delta x[3a]={1.07~r_s}$, and HR with $\Delta x[3a]={0.66~r_s}$.}
    \label{fig:Torque_res}
\end{figure}

In Fig.~\ref{fig:Torque_res}, we show the torque values $\bar \xi$ time-averaged over 100 initial binary orbits, or equivalently in the SMA range between $54.5~r_s$ and $52.6~r_s$, for three different resolutions of the gaseous disk. We quantify the resolution by measuring an equivalent inter-particle spacing $\Delta x[3a]$ evaluated at $R=3a$. We name the three simulations as: low-resolution (LR) with $\Delta x[3a]={1.31~r_s}$, mid-resolution (MR) with $\Delta x[3a]={1.07~r_s}$, and high-resolution (HR) with $\Delta x[3a]={0.66~r_s}$. Fig.~\ref{fig:Torque_res} shows the value of $\bar \xi$ for the three resolutions.
Since the values measured from our MR and HR runs are very similar, within $0.4\%$, we can conclude that the MR run is sufficiently converged and we therefore further evolve it until $278$ orbits to measure the dephasing induced by the interaction of the binary with the disk. We run the {gas+PN2} run at the MR resolution of $\Delta x[3a]={1.07~r_s}$ for the same elapsed physical time in order to measure the differences between the two simulations (see Section \S~\ref{Sec:Results}). Since the {gas+PN2} MR setup has no fast inspiral due to GWs, we can reasonably assume it is also converged.

\end{document}